\documentclass[12pt]{article}
\usepackage{amsmath}
\usepackage{latexsym}
\usepackage{bbm}
\usepackage[T1]{fontenc}
\usepackage[latin1]{inputenc}
\usepackage{geometry}

\def\g{\gamma}

\def\d{\delta}

\def\e{\varepsilon}

\def\i{\iota}

\def\l{\lambda}
\def\m{\mu}
\def\n{\nu}

\def\r{\rho}

\def\s{\sigma}
\def\t{\tau}

\def\G{\Gamma}
\def\D{\Delta}

\def\L{\Lambda}

\def\pa{\partial}

\def\half{\frac{1}{2}}

\def\IC{\mathbbm C}

\def\IR{\mathbbm R}
 
\def\IZ{\mathbbm Z}

\def\half{\frac{1}{2}}

\def\tr{{\rm tr}}

\def\and{{\rm and}}

\def\ie{{\it i.e.,} }

\begin{document}
\vspace*{-.6in} \thispagestyle{empty}
\begin{flushright}
CALT-68-2693
\end{flushright}
\baselineskip = 18pt

\vspace{1.5in} {\Large
\begin{center}
\bf Studies of the ABJM Theory in a \\Formulation with Manifest
SU(4) R-Symmetry\end{center}}

\begin{center}
Miguel A. Bandres, Arthur E. Lipstein and John H. Schwarz
\\
\emph{California Institute of Technology\\ Pasadena, CA  91125, USA}
\end{center}
\vspace{1in}

\begin{center}
\textbf{Abstract}
\end{center}
\begin{quotation}
\noindent We examine the three-dimensional ${\cal N} =6$
superconformal Chern--Simons theory with $U(N) \times U(N)$ gauge
symmetry, which was recently constructed by Aharony, Bergman,
Jafferis, and Maldacena (ABJM). Using a formulation with manifest
$SU(4)$ R-symmetry and no auxiliary fields, we verify in complete
detail both the Poincar\'e supersymmetry and the conformal
supersymmetry of the action. Together, these imply the complete
$OSp(6|4)$ superconformal symmetry of the theory. The potential,
which is sixth order in scalar fields, is recast as a sum of
squares.
\end{quotation}

\newpage

\pagenumbering{arabic}

\section{Introduction}

This paper examines a new class of superconformal field theories in
three dimensions that was recently discovered by Aharony, Bergman,
Jafferis, and Maldacena (ABJM) \cite{Aharony:2008ug}. These theories
are superconformal Chern--Simons gauge theories with ${\cal N} =6$
supersymmetry. When the gauge group is chosen to be $U(N) \times
U(N)$ and the Chern--Simons level is $k$, these theories are
conjectured to be dual to M-theory on $AdS_4 \times S^7/\IZ_k$ with
$N$ units of flux. More precisely, this is the appropriate dual
description for $N^{1/5} >> k$. In the opposite limit, $N^{1/5} << k
<<N$, a dual description in terms of type IIA string theory on
$AdS_4 \times \IC P^3$ is more appropriate. A large-$N$ expansion
for fixed 't Hooft parameter $\l =N/k$ can be defined. These
developments raise the hope that this duality can be analyzed in the
same level of detail as has been done for the duality between ${\cal
N} =4$ super Yang--Mills theory with a $U(N)$ gauge group in four
dimensions and type IIB superstring theory on $AdS_5 \times S^5$
with $N$ units of flux.

Even though the ABJM paper is very recent, quite a few papers have
already appeared that examine various of its properties as well as
possible generalizations. Among the first are
\cite{Benna:2008zy}--\cite{Grignani:2008te}. New superconformal
Chern--Simons theories with ${\cal N} = 5$ supersymmetry have been
constructed in \cite{Hosomichi:2008jb}. (This paper also does many
other things.) Certain of these ${\cal N} = 5$ theories should be
dual to the $D_{k+2}$ orbifolds described in \cite{Hanany:2008qc}.
Also, in a very interesting recent paper \cite{Bagger:2008se},
Bagger and Lambert show that the ABJM theories correspond to a class
of 3-algebras in which the bracket $[T^a, T^b, T^c]$ is no longer
antisymmetric in all three indices. The actions and supersymmetry
transformations that are derived in
\cite{Hosomichi:2008jb,Bagger:2008se} appear to be equivalent to the
actions and supersymmetry transformations that are obtained in this
paper (without reference to 3-algebras).

The three-dimensional superconformal field theories of coincident
M2-branes were initially defined as infrared fixed points of super
Yang--Mills theories, \ie as limits of the form $g_{\rm YM} \to
\infty$. In \cite{Schwarz:2004yj} it was proposed that these fixed
points could be reformulated in a more useful dual formulation
analogous to a Seiberg dual. It was suggested that the theory would
be a gauge theory in which the gauge fields couple to dimension-1/2
scalar and dimension-1 spinor fields. Since all terms should be
dimension 3, there should be no dimension-4 $F^2$ kinetic terms, but
dimension-3 Chern--Simons terms would be allowed. An approach to
constructing such theories based on considering multiple M2-branes
ending on an M5-brane was proposed in \cite{Basu:2004ed}. Several
years later, a specific example of such a superconformal
Chern--Simons theory with maximal (${\cal N} =8$) supersymmetry was
constructed by Bagger and Lambert
\cite{Bagger:2006sk,Bagger:2007jr,Bagger:2007vi} and by Gustavsson
\cite{Gustavsson:2007vu,Gustavsson:2008dy}. This theory is parity
conserving and has $SO(4) = SU(2) \times SU(2)$ gauge symmetry
\cite{Bandres:2008vf,VanRaamsdonk:2008ft}. The scalars and spinors
are 4-vectors of $SO(4)$, or (equivalently) bifundamentals of $SU(2)
\times SU(2)$.

The BLG theory was conjectured \cite{Bandres:2008vf} and proved
\cite{Papadopoulos:2008sk,Gauntlett:2008uf} to be the unique theory
of this type with maximal supersymmetry. (Generalizations based on
Lorentzian 3-algebras \cite{Gomis:2008uv,Benvenuti:2008bt,Ho:2008ei}
turned out to be equivalent to the original super Yang--Mills
theories once the ghosts were eliminated
\cite{Bandres:2008kj,Gomis:2008be,Ezhuthachan:2008ch}.) This left
the possibility of considering theories with reduced supersymmetry.
A large class of superconformal Chern--Simons theories with ${\cal
N} =4$ supersymmetry was constructed by Gaiotto and Witten
\cite{Gaiotto:2008sd}. This was generalized to include twisted
hypermultiplets in \cite{Hosomichi:2008jb,Hosomichi:2008jd}. This
generalization includes the Bagger--Lambert theory as a special
case. Moreover, all the ABJM theories turn out to be special cases
of the generalized Gaiotto--Witten theories in which the
supersymmetry is enhanced to ${\cal N} =6$. The dual M-theory
picture requires that for levels $k=1,2$ the ABJM theories should
have ${\cal N} =8$ supersymmetry. However, this has not yet been
demonstrated explicitly.

The purpose of this paper is to recast the ABJM theory in a form for
which the $SU(4)$ R-symmetry of the action and the supersymmetry
transformations is manifest and to use this form to study some of
its properties. The existence of such formulas is a consequence of
what was found in \cite{Aharony:2008ug}. We also verify the
conformal supersymmetry of the action, which is not a logical
consequence of previous results. Since this symmetry is a necessary
requirement for the validity of the proposed duality, its
verification can be viewed as an important and nontrivial test of
the duality. We also recast the potential, which is sixth order in
the scalar fields, in a new form.\footnote{A similar formula also
appears in \cite{Bagger:2008se}.} This new form should be useful for
studying the moduli space of supersymmetric vacua of the theory, as
well as the vacuum structure of various deformations of the ABJM
theory. Although we discuss the gauge group $U(N) \times U(N)$, all
of our analysis also holds for the straightforward generalization to
$U(M) \times U(N)$.

Some of our results are new and others confirm results that have
been obtained previously. The ABJM theories were formulated in
\cite{Aharony:2008ug} using auxiliary fields associated with ${\cal
N} =2$ superfields. In this formulation only an $SU(2) \times SU(2)$
subgroup of the $SU(4)$ R-symmetry is manifest, though the full
$SU(4)$ symmetry has been deduced. In addition,
\cite{Aharony:2008ug} deduced a manifestly $SU(4)$ invariant form of
the scalar field potential, which is sixth order in the scalar
fields. The quartic interaction terms that have two scalar and two
spinor fields were also recast in an $SU(4)$ covariant form in
\cite{Benna:2008zy}. Our results are in agreement with both of
these.

\section{The $U(1)\times U(1)$ Theory}

The field content of ABJM theories consists of scalars, spinors, and
gauge fields. The $U(1)\times U(1)$ theory has fewer indices to keep
track of, and it is quite a bit simpler, than the full $U(N)\times
U(N)$ theory; so it is a good place to start.

There are four complex scalars $X_A$ and their adjoints $X^A$. (We
choose not to use adjoint or complex conjugation symbols to keep the
notation from becoming too cumbersome.)  A lower index labels the
${\bf 4}$ representation of the global $SU(4)$ R-symmetry and an
upper index labels the complex-conjugate ${\bf \bar 4}$
representation.

Similarly, the fermi fields are $\Psi^A$ and $\Psi_A$. These are
also two-component spinors, though that index is not displayed. As
usual, the notation $\bar\Psi^A$ or $\bar\Psi_A$ implies transposing
the spinor index and right multiplication by $\g^0$. Note, however,
that for our definition there is no additional complex conjugation,
so in all cases a lower index indicates a ${\bf 4}$ and an upper
index indicates a ${\bf \bar 4}$. With these conventions various
identities that hold for Majorana spinors can be used for these
spinors, as well, even though they are complex (Dirac). For example,
$\bar\Psi^A \Psi_B = \bar\Psi_B \Psi^A$. The $2\times 2$ Dirac
matrices satisfy $\{\g^\m,\g^\n\} = 2\eta^{\m\n}$. The index $\m
=0,1,2$ is a 3-dimensional Lorentz index, and the signature is
$(-,+,+)$. It is convenient to use a Majorana representation, which
implies that $\g^\m$ is real. We also choose a representation for
which $\g^{\m\n\l} = \e^{\m\n\l}$. In particular, this means that
$\g^0 \g^1 \g^2 =1$. For example, one could choose $\g^0 = i\s^2$,
$\g^1 =\s^1$, and $\g^2 =\s^3$.

The $U(1)$ gauge fields are denoted $A_\m$ and $\hat A_\m$. The
fields $X_A$ and $\Psi^A$ have $U(1)$ charges $(+,-)$, while their
adjoints have charges $(-,+)$. Thus, for example,
\begin{equation}
D_\m X_A = \pa_\m X_A  +i (A_\m - \hat A_\m) X_A.
\end{equation}
and
\begin{equation}
D_\m X^A = \pa_\m X^A  - i (A_\m - \hat A_\m) X^A.
\end{equation}

We choose to normalize fields so that the level-$k$ Lagrangian is
$k$ times the level-1 Lagrangian. With this convention, the $N=1$
action is
\begin{equation}
S = \frac{k}{2\pi} \int d^3 x \left( - D^\m X^A D_\m X_A
+i\bar\Psi_A \g^\m D_\m \Psi^A + \half \e^{\m\n\l} (A_\m \pa_\n A_\l
- \hat A_\m \pa_\n \hat A_\l) \right).
\end{equation}
The claim is that this action describes an ${\cal N} =6$
superconformal theory with $OSp(6|4)$ superconformal symmetry. The
R-symmetry is $Spin(6)=SU(4)$ and the conformal symmetry is $Sp(4) =
Spin (3,2)$. The supercharges transform as the ${\bf 6}$
representation of $SU(4)$. Both the Poincar\'e and conformal
supercharges are 6-vectors. Each accounts for 12 of the 24 fermionic
generators of the superconformal algebra.

The antisymmetric product of two ${\bf 4}$s gives a ${\bf 6}$. The
invariant tensor (or Clebsch--Gordan coefficients) describing this
is denoted $\G^I_{AB} = -\G^I_{BA}$, since these can be interpreted
as six matrices satisfying a Clifford algebra. More precisely, if
one also defines $\tilde\G^I = (\G^I)^\dagger$, or in components
\begin{equation}
\tilde\G^{IAB} = \half \e^{ABCD} \G^I_{CD} =
-\left(\G^I_{AB}\right)^*,
\end{equation}
then\footnote{An explicit realization in terms of Pauli matrices is
given by $\G^1 = i\s_2\otimes1$, $\G^2 = \s_2\otimes\s_1$, $\G^3 =
\s_2\otimes\s_3$, $\G^4 = 1\otimes\s_2$, $\G^5 = i\s_1\otimes\s_2$,
$\G^6 = i\s_3\otimes\s_2$.}
\begin{equation}
\G^I \tilde\G^J + \G^J \tilde\G^I = 2\d^{IJ}.
\end{equation}
Note that $\g^\m$ are $2\times2$ matrices and $\G^I$ are $4 \times
4$ matrices. They act on different vector spaces, and therefore they
trivially commute with one another.

The supersymmetry transformations of the matter fields are
\begin{equation}
\d X_A = i \G^I_{AB} \bar\Psi^B \e^I
\end{equation}
\begin{equation}
\d \Psi_A =  \G^I_{AB}\g^\m \e^I D_\m X^B
\end{equation}
and their adjoints, which are
\begin{equation}
\d X^A = -i \tilde\G^{IAB} \bar\Psi_B \e^I
\end{equation}
\begin{equation}
\d \Psi^A = - \tilde\G^{IAB}\g^\m \e^I D_\m  X_B .
\end{equation}
For the gauge fields we have
\begin{equation} \label{deltaA}
\d A_\m = \d \hat A_\m = -\G^I_{AB} \bar\Psi^A \g_\m \e^I  X^B -
\tilde\G^{IAB} \bar\Psi_A \g_\m \e^I  X_B .
\end{equation}
The verification that these leave the action invariant is given in
the Appendix.

Note that the covariant derivatives only involve $A_-$, where
\begin{equation}
A_\pm = A \pm \hat A.
\end{equation}
Therefore, let us rewrite the Chern--Simons terms using
\cite{Cattaneo:1995tw}
\begin{equation}
\int (A\wedge dA - \hat A\wedge d \hat A) = \int A_+\wedge d A_- =
\int A_- \wedge d A_+.
\end{equation}
Since this is the only appearance of $A_+$ in the action, it can be
integrated out to give the delta functional constraint
\begin{equation}
F_- = dA_- =0.
\end{equation}
The $A_-$ equation of motion, on the other hand, just identifies
$F_+$ with the dual of the charge current. Since the kinetic terms
are defined with a flat connection $A_-$, this is just a free theory
when the topology is trivial, which is the case for $k=1$. Then this
theory has ${\cal N} =8$ superconformal symmetry.

ABJM proposes to treat $F_+$ as an independent variable and to add a
Lagrange multiplier term to ensure that $F_+$ is a curl
\begin{equation}
S_\t = \frac{1}{4\pi} \int \t \e^{\m\n\l} \pa_\m F_{+\n\l} d^3 x.
\end{equation}
Then the quantization condition on $F_+$ requires that $\t$ has
period $2\pi$. They then explain that after gauge fixing $\t =0$ one
is left with a residual $\IZ_k$ gauge symmetry under which $X^A \to
{\rm exp}(2\pi i/k)X^A$ and similarly for $\Psi_A$. Thus one is left
with a sigma model on $\IC^4/\IZ_k$. This breaks the supersymmetry
from ${\cal N}=8$ to ${\cal N}=6$ for $k>2$. The reason for this is
that the 8-component $Spin(8)$ supercharge decomposes with respect
to the $SU(4) \times U(1)$ subgroup as $6_0 + 1_2 + 1_{-2}$. Because
of their $U(1)$ charges, the singlets transform under a $\IZ_k$
transformation as $Q \to {\rm exp}(\pm 4\pi i/k)Q$. Therefore two of
the supersymmetries are broken for $k>2$.

This analysis of the $U(1)$ factors continues to apply in the
$U(N)\times U(N)$ theories with $N>1$. The Bagger--Lambert theory
corresponds to the gauge group $SU(2)\times SU(2)$. Since it has no
$U(1)$ factors, no discrete $Z_k$ gauge symmetry arises, and this
theory has ${\cal N} =8$ superconformal symmetry for all values of
$k$. So, it is different from the $U(2)\times U(2)$ ABJM theory, and
its interpretation in terms of branes or geometry (see
\cite{Lambert:2008et,Distler:2008mk}) must also be different.

\section{The $U(N)\times U(N)$ Theory}

The field content of the $U(N)\times U(N)$ ABJM theory consists of
four $N \times N$ matrices of complex scalars $(X_A)^a{}_{\hat a}$
and their adjoints $(X^A)^{\hat a}{}_a$. These transform as $({\bf
\bar N}, {\bf N})$ and $({\bf N}, {\bf \bar N})$ representations of
the gauge group, respectively.  Similarly, the spinor fields are
matrices $(\Psi^A)^a{}_{\hat a}$ and their adjoints $(\Psi_A)^{\hat
a}{}_a$. The $U(N)$ gauge fields are hermitian matrices $A^a{}_b$
and $\hat A^{\hat a}{}_{\hat b}$. In matrix notation, the covariant
derivatives are
\begin{equation}
D_\m X_A = \pa_\m X_A  +i (A_\m X_A - X_A\hat A_\m)
\end{equation}
and
\begin{equation}
D_\m X^A = \pa_\m X^A  + i( \hat A_\m X^A - X^A A_\m)
\end{equation}
with similar formulas for the spinors. Infinitesimal gauge
transformations are given by
\begin{equation}
\d A_\m = D_\m \L = \pa_\m \L+i [A_\m,\L],
\end{equation}
\begin{equation}
\d \hat A_\m = D_\m \hat\L = \pa_\m\hat\L +i [\hat A_\m,\hat\L],
\end{equation}
\begin{equation}
\d X_A =-i\L X_A +i X_A \hat\L,
\end{equation}
and so forth.

The action consists of terms that are straightforward
generalizations of those of the $U(1) \times U(1)$ theory, as well
as new interaction terms that vanish for $N=1$. The kinetic and
Chern--Simons terms are
\begin{equation}
S_{\rm kin} = \frac{k}{2\pi} \int d^3 x \,\tr\left( - D^\m X^A D_\m
X_A +i\bar\Psi_A \g^\m D_\m \Psi^A\right).
\end{equation}
and
\begin{equation}
S_{\rm CS} = \frac{k}{2\pi} \int d^3 x  \, \e^{\m\n\l} \tr\Big(
\half A_\m \pa_\n A_\l  + \frac{i}{3} A_\m A_\n A_\l- \half \hat
A_\m \pa_\n \hat A_\l - \frac{i}{3} \hat A_\m \hat A_\n \hat
A_\l\Big) .
\end{equation}
Additional interaction terms of the schematic form $X^2\Psi^2$ and
$X^6$ remain to be determined. These terms are not required to
deduce the equations of motion of the gauge fields, which are
\begin{equation}
J^\m = \half\e^{\m\n\l} F_{\n\l} \quad \and \quad \hat J^\m =
-\half\e^{\m\n\l} \hat F_{\n\l},
\end{equation}
where
\begin{equation}
J^\m = iX_A D^\m X^A -i D^\m X_A X^A - \bar\Psi^A \g^\m \Psi_A
\end{equation}
and
\begin{equation}
\hat J^\m = iX^A D^\m X_A -i D^\m X^A X_A - \bar\Psi_A \g^\m \Psi^A.
\end{equation}
Note that in the special case of $U(1) \times U(1)$ one has $J^\m =
-\hat J^\m$, and hence the equations of motion imply $F_{\m\n} =
\hat F_{\m\n}$.

In matrix notation, the supersymmetry transformations of the matter
fields are
\begin{equation}
\d X_A = i \G^I_{AB} \bar\e^I  \Psi^B
\end{equation}
and
\begin{equation}
\d \bar\Psi_A =  -\G^I_{AB}\bar\e^I\g^\m D_\m X^B + \d_3 \bar\Psi_A
\end{equation}
or equivalently
\begin{equation}
\d \Psi_A =  \G^{I}_{AB}\g^\m \e^I D_\m  X^B + \d_3 \Psi_A.
\end{equation}
and their adjoints, which are
\begin{equation}
\d X^A = -i \tilde\G^{IAB} \bar\Psi_B \e^I
\end{equation}
and
\begin{equation}
\d \Psi^A =  -\tilde\G^{IAB}\g^\m \e^I D_\m  X_B + \d_3 \Psi^A.
\end{equation}
or equivalently
\begin{equation}
\d \bar\Psi^A =  \tilde\G^{IAB}\bar\e^I\g^\m D_\m X_B + \d_3
\bar\Psi^A.
\end{equation}
The terms denoted $\d_3$ are cubic in $X$ and are given below. The
supersymmetry transformations of the gauge fields are
\begin{equation}
\d A_\m =  \G^I_{AB}\bar\e^I \g_\m \Psi^A  X^B - \tilde\G^{IAB}
X_B\bar\Psi_A \g_\m \e^I
\end{equation}
\begin{equation}
\d \hat A_\m =  \G^I_{AB} X^B \bar\e^I \g_\m \Psi^A - \tilde\G^{IAB}
\bar\Psi_A \g_\m \e^I  X_B.
\end{equation}
Note that $\d A_\m \neq \d \hat A_\m$ for $N>1$. They are matrices
in different spaces.

In the Appendix we show that supersymmetry requires the choice
\begin{equation}
\d_3 \Psi^A = N^{IA} \e^I \quad \and \quad \d_3 \Psi_A = N^I_A\e^I,
\end{equation}
where
\begin{equation} \label{NsupA}
N^{IA} = \tilde\G^{IAB} (X_C X^C X_B - X_B X^C X_C) -2
\tilde\G^{IBC} X_B X^A X_C.
\end{equation}
and
\begin{equation} \label{NsubA}
N^I_A = (N^{IA})^{\dagger} = \G^I_{AB} (X^C X_C X^B - X^B X_C X^C)
-2 \G^I_{BC} X^B X_A X^C.
\end{equation}
Note that these expressions vanish when the matrices $X^A$ (and
their adjoints $X_A$) are diagonal.

All the possible structures for the $\Psi^2 X^2$ terms are
\begin{equation}
L_{\rm 4a} =   i\e^{ABCD} \tr(\bar\Psi_A  X_B \Psi_C  X_D)
-i\e_{ABCD}\tr( \bar\Psi^A X^B \Psi^C X^D )
\end{equation}
\begin{equation}
L_{\rm 4b} = i \tr( \bar\Psi^A \Psi_A X_B X^B) - i\tr( \bar\Psi_A
\Psi^A X^B X_B)
\end{equation}
\begin{equation}
L_{\rm 4c} = 2i\tr( \bar\Psi_A \Psi^B X^A X_B) - 2i\tr( \bar\Psi^B
\Psi_A X_B X^A)
\end{equation}
The coefficients are chosen so that $L_4 = L_{\rm 4a} + L_{\rm 4b} +
L_{\rm 4c}$ is the correct result required by supersymmetry, as is
demonstrated in the Appendix.

The lagrangian also contains a term $L_6 =-V$ that is sixth order in
the scalar fields. The scalar potential $V$ is expected to be
nonnegative and to vanish for a supersymmetric vacuum. An $SU(4)$
covariant formula for $V$ in terms of the fields $X^A$ and $X_A$ has
been given in \cite{Aharony:2008ug,Benna:2008zy}
\[
V = -\frac{1}{3}\tr\Big[X^{A}X_{A}X^{B}X_{B}X^{C}X_{C}
+X_{A}X^{A}X_{B}X^{B}X_{C}X^{C}
\]
\begin{equation}\label{potential}
+4X_{A}X^{B}X_{C}X^{A}X_{B}X^{C}
-6X^{A}X_{B}X^{B}X_{A}X^{C}X_{C}\Big] ,
\end{equation}
a result that we confirm in the Appendix.

This formula for $V$ is not expressed as a sum of squares, which
makes it inconvenient for determining the extrema. For a
supersymmetric vacuum, $\d \Psi^A = \d \Psi_A =0$. In particular,
for a solution in which the scalar fields $X^A$ and $X_A$ are
constant, and the gauge fields vanish, the variations $\d_3 \Psi^A$
and $\d_3 \Psi_A$ should vanish. This implies that $N^{IA}=0$ and
$N^I_A = (N^{IA})^{\dagger} =0$. The way to ensure these
requirements, as well as manifest $SU(4)$ symmetry, is for the
potential to take the form
\begin{equation} \label{potential2}
V = \frac{1}{6} \tr(N^{IA} N^I_A) .
\end{equation}
The definitions of $N^{IA}$ and $N^I_A$ are given in
Eqs.~(\ref{NsupA}) and (\ref{NsubA}). It is straightforward to
verify the equivalence of Eqs.~(\ref{potential}) and
(\ref{potential2}) for this choice of the coefficient by using the
key identity
\begin{equation}
\Gamma_{AB}^{I}\tilde{\Gamma}^{ICD} = -2\delta_{AB}^{CD}.
\end{equation}
The indicated relationship between the potential and $\d_3\Psi$ in
Eq.~(\ref{potential2}) should be quite general in theories of this
type. As has already been noted, $N^{IA}$ and $N^I_A$ vanish when
the scalar fields are diagonal matrices. To get the expected moduli
space, these should be the only choices for which they vanish
(modulo gauge transformations).

\section{Conclusion}

The study of ABJM theories has become a hot topic. The technology
that has been developed in the study of the duality between
four-dimensional superconformal gauge theories and $AdS_5$ vacua of
type IIB superstring theory can now be adapted to a new setting. It
should now be possible to study the duality between
three-dimensional superconformal Chern--Simon theories and $AdS_4$
vacua of type IIA superstring theory and M-theory. A great deal
should be learned in the process, and there may even be applications
to other areas of physics.

Our contribution to this subject is modest: We have verified the
Poincar\'e supersymmetries of the ABJM theory in a formalism with
manifest $SU(4)$ symmetry. The action that we obtained agrees with
results given in \cite{Aharony:2008ug,Benna:2008zy,Bagger:2008se}.
We have also verified by explicit calculation that this action has
the conformal supersymmetries that are required by the proposed
duality. Since this is not implied by any previous calculations, it
is an important (and nontrivial) test of the duality. Taken together
with the Poincar\'e supersymmetries, this implies the full
$OSp(6|4)$ superconformal symmetry of the action. We have also
recast the sextic potential as a sum of squares in
Eq.~(\ref{potential2}), a form that should prove useful in future
studies.

\section*{Acknowledgments}

This work was supported in part by the U.S. Dept. of Energy under
Grant No. DE-FG03-92-ER40701. MAB acknowledges support from the
Secretar\'{\i}a de Educaci\'{o}n Publica de M\'{e}xico. JHS wishes
to acknowledge the hospitality of the Aspen Center for Physics. He
also acknowledges helpful discussions with Igor Klebanov.

\section*{Appendix: Verification of Superconformal Symmetry}

\subsection*{The $U(1)\times U(1)$ Theory}

Let us check the supersymmetry of the $U(1)\times U(1)$ theory.  We
only analyze half of the terms, since the other half are just their
adjoints. Omitting the factor of $k/2\pi$, the variation of the
Lagrangian contains (dropping total derivatives)
\begin{equation}
\D_1 = -D^\m X^A D_\m \d X_A =iD^2 X^A \bar\e^I\G^I_{AB} \Psi^B
\end{equation}
and
\newpage
$$
\D_2 = i\d\bar\Psi_A \g \cdot D \Psi^A = -i\G^I_{AB}\bar\e^I\g \cdot
D X^B \g \cdot D \Psi^A
$$
\begin{equation}
= i\G^I_{AB}\bar\e^I D^2 X^B  \Psi^A -
\half\G^I_{AB}\bar\e^I\g^{\r\m} (F_{\r\m} -\hat F_{\r\m})X^B \Psi^A
.
\end{equation}
Note that the gauge fields only appear in the covariant derivatives
in the combination $ A - \hat A$, which has a vanishing
supersymmetry variation. The variation of the Chern--Simons term,
using the first term in Eq.~(\ref{deltaA}), contributes
\begin{equation}
\D_3 =  \half\e^{\m\n\l} \bar\e^I \g_\m \Psi^A \G^I_{AB} X^B (
F_{\n\l} - \hat F_{\n\l}).
\end{equation}
Using $\e^{\m\n\l} \g_\m = \g^{\n\l}$, we see that $\D_1 + \D_2 +
\D_3 =0$. The other half of the terms in the variation of the
action, which are the adjoints of the ones considered here, cancel
in the same way. The conserved supersymmetry current can be computed
by the standard Noether procedure. This gives (aside from an
arbitrary normalization)
\begin{equation}
Q_\m^I = \G^I_{AB} \g\cdot D X^A  \g_\m \Psi^B - \tilde\G^{IAB}
\g\cdot D X_A \g_\m \Psi_B.
\end{equation}
One can check this result by computing the divergence. This vanishes
as a consequence of the equations of motion $\g\cdot D\Psi^B =0$,
$D\cdot D X^A =0$, and $F_{\m\n} -\hat F_{\m\n}=0$.

Let us now explore the conformal supersymmetry, with an
infinitesimal spinor parameter $\eta^I$, using the method explained
in \cite{Bandres:2008vf}. As a first try, consider replacing $\e^I$
by $\g\cdot x \eta^I$ in the preceding equations, since this has the
correct dimensions. Using $\pa_\m\e(x) = \g_\m \eta$ and $\g^\m
\g^\rho \g_\m = - \g^\rho$, this gives a variation of the action
that almost cancels, except for a couple of terms.  These remaining
terms can be canceled by including an additional variation of the
spinor fields. It has the form
\begin{equation}
\d'\Psi^A = -\tilde\G^{IAB} \eta^I X_B \quad \and \quad \d' \Psi_A =
\G^I_{AB} \eta^I X^B.
\end{equation}
Correspondingly, the conserved superconformal current is
\begin{equation}
S_\m^I = \g\cdot x \, Q_\m^I +\G^I_{AB} X^A \g_\m \Psi^B -
\tilde\G^{IAB} X_A \g_\m \Psi_B.
\end{equation}
As a check, one can compute the divergence using the conservation of
$Q^I_\m$ and the spinor field equation of motion
\begin{equation}
\pa^\m S_\m^I =\g^\m Q^I_\m + \G^I_{AB} \g\cdot D X^A  \Psi^B -
\tilde\G^{IAB} \g \cdot D X_A  \Psi_B = 0.
\end{equation}

The various bosonic $OSp(6|4)$ symmetry transformations are obtained
by commuting $\e$ and $\eta$ transformations. Of these only the
conformal transformation, obtained as the commutator of two $\eta$
transformations, is not a manifest symmetry of the action. It is
often true that scale invariance implies conformal symmetry.
However, this is not a general theorem, so it is a good idea to
check the conformal symmetry (or the conformal supersymmetry)
explicitly.

\subsection*{The $U(N)\times U(N)$ Theory}

Let us now examine the supersymmetry of the $U(N)\times U(N)$
theory. Some of the terms are simple generalizations of those
examined in the $N=1$ case and will not be described here. Rather,
we focus on those that only arise for $N>1$. We will first determine
the quartic $\Psi^2 X^2$ term (called $L_4$) in the action by
requiring that the variation of its $X$ fields cancels the terms
that arise from varying the gauge fields in the spinor kinetic term.
Since these terms are cubic in $\Psi$, various Fierz identities are
required. The second step is to determine the variation $\d_3 \Psi$
by requiring that this variation of the spinor kinetic term cancels
against the lowest-order variation of the $\Psi$ fields in $L_4$ and
the variation of the gauge fields in the scalar kinetic term. The
third and final step is to determine $L_6$ by arranging that its
variation cancels against the $\d_3 \Psi$ variation of $L_4$. After
this has been completed, we verify the conformal supersymmetry.

\subsubsection*{Determination of $L_4$}

A useful identity involving four two-component Majorana spinors,
obtained by a Fierz transformation, is
\begin{equation}
\bar\psi_1 \g_\m \psi_2 \bar\psi_3 \g^\m \e = -2 \bar\e \psi_1
\bar\psi_2 \psi_3 - \bar\psi_1 \psi_2 \bar\e \psi_3.
\end{equation}
Juggling the indices this can be recast in the form
\begin{equation}
\bar\e \g_\m \psi_1 \bar\psi_2 \g^\m \psi_3 = -2 \bar\psi_1 \psi_2
\bar\e \psi_3 - \bar\e \psi_1 \bar\psi_2  \psi_3.
\end{equation}
These will be useful for eliminating Dirac matrices from equations
that arise later. As written, these relations preserve the 123
sequence of the spinors, which is convenient if they are matrices
that are to be multiplied. However, the right-hand sides can be
rewritten  in other ways without Dirac matrices using the relation
\begin{equation} \label{psi123}
\psi_1 \bar\psi_2 \psi_3 + \psi_2 \bar\psi_3 \psi_1 + \psi_3
\bar\psi_1 \psi_2 =0.
\end{equation}
This equation will also be useful.

Varying the gauge fields in the spinor kinetic term of the
$U(N)\times U(N)$ theory (dropping a factor of $k/2\pi$) gives
\begin{equation}
\tr\left( \bar\Psi_A \g^\m (-\d A_\m \Psi^A + \Psi^A \d \hat
A_\m)\right).
\end{equation}
Keeping only the terms with two superscripts on spinor fields, since
the other terms are just their adjoints, leaves
\begin{equation}
\G^I_{BC} \tr(-\bar\Psi^A\g^\m \Psi_A \bar\Psi^B \g_\m \e^I X^C
+\bar\e^I \g^\m \Psi^B \bar\Psi_A \g_\m \Psi^A X^C).
\end{equation}
Inserting the identities above, so as to eliminate Dirac matrices
while retaining the order of the matrices, which are implicitly
multiplied, leaves
$$
\G^I_{BC} \tr\Big(2\bar\e^I\Psi^A \bar\Psi_A \Psi^B X^C
+\bar\Psi^A\Psi_A \bar\e^I\Psi^B X^C -2\bar\Psi^B \Psi_A
\bar\e^I\Psi^A X^C -\bar\e^I \Psi^B \bar\Psi_A \Psi^A X^C\Big)
$$
\begin{equation}
= i\tr(\bar\Psi^A\Psi_A \d X_B X^B) -i\tr(\bar\Psi_A\Psi^A X^B \d
X_B)
\end{equation}
$$
+2 \G^I_{BC}\tr(\bar\e^I\Psi^A[\bar\Psi_A\Psi^B X^C -X^C \bar\Psi^B
\Psi_A]).
$$

Now consider varying the $X$ fields in the second term in $L_{4a}$.
This gives
$$
-2i\e_{ABCD}\tr( \bar\Psi^A \d X^B \Psi^C X^D ) =
-2\tilde\G^{IBE}\e_{ABCD}\tr( \bar\Psi^A \bar\e^I\Psi_E \Psi^C X^D )
$$
$$
=-\e^{BEFG}\e_{ABCD}\G^I_{FG}\tr( \bar\Psi^A \bar\e^I\Psi_E \Psi^C
X^D )
$$
$$
=  \d^{EFG}_{ACD}\G^I_{FG}\tr( \bar\Psi^A \bar\e^I\Psi_E \Psi^C X^D)
$$
\begin{equation}
= - \d^{EFG}_{ACD}\G^I_{FG}\tr( \bar\Psi^A\Psi_E\bar\e^I\Psi^C X^D+
\bar\Psi^A\e^I\bar\Psi_E\Psi^CX^D)
\end{equation}
$$
= -2i\tr(\bar\Psi^A\Psi_A \d X_B X^B) + 2i\tr(\bar\Psi_A\Psi^A  X^B
\d X_B) +2i\tr(\bar\Psi^A\Psi_B \d X_A X^B)
$$
$$
-2i\tr(\bar\Psi_A\Psi^B  X^A \d X_B) -2
\G^I_{BC}\tr(\bar\e^I\Psi^A[\bar\Psi_A\Psi^B X^C -X^C \bar\Psi^B
\Psi_A]),
$$
where we have used Eq.~(\ref{psi123}). Here we have used the
definition
\begin{equation}
\d^{DEF}_{ABC} = 6\d^{[D}_A \d^E_B \d^{F]}_C.
\end{equation}
These two sets of terms combine to leave
$$
-i\tr(\bar\Psi^A\Psi_A \d X_B X^B) + i\tr(\bar\Psi_A\Psi^A  X^B \d
X_B)
$$
\begin{equation}
+2i\tr(\bar\Psi^B\Psi_A \d X_B X^A) -2i\tr(\bar\Psi_A\Psi^B X^A \d
X_B).
\end{equation}
These terms are canceled in turn by varying $X_B$ in $L_{\rm 4b}$
and $L_{\rm 4c}$. Thus, terms of this structure in the supersymmetry
transformations cancel for the choice of $L_4$ given in section 3.
The adjoint terms cancel in the same way.

Since we now have the complete dependence of the action on spinor
fields, we can deduce the spinor field equations of motion. They are
\[
\g\cdot D\Psi^A =  - 2\e^{ABCD}  X_B \Psi_C  X_D  -  X_B X^B \Psi^A
+\Psi^A X^B X_B
\]
\begin{equation} \label{PsiEOMa}
- 2 \Psi^B X^A X_B + 2 X_B X^A \Psi^B
\end{equation}
and its adjoint
\[
\g\cdot D\Psi_A =   2\e_{ABCD}  X^B \Psi^C  X^D  +  X^B X_B \Psi_A
-\Psi_A X_B X^B
\]
\begin{equation} \label{PsiEOMb}
+ 2 \Psi_B X_A X^B - 2 X^B X_A \Psi_B.
\end{equation}

\subsubsection*{Determination of $\d_3 \Psi$}

Having determined $L_4$, we are now in a position to determine
$\d_3\Psi$ by computing terms of the schematic structure $\tr(\Psi_A
DX_B X^C X_D)$, $\tr(\Psi_A X_B DX^C X_D)$, and $\tr(\Psi_A X_B X^C
DX_D)$ that arise from varying the gauge fields in the $X$ kinetic
term and varying the spinor fields in $L_4$. The adjoint terms work
the same way. The terms of the indicated structure that arise from
varying the gauge fields in the $X$ kinetic term are
\begin{equation}
i\tilde\G^{IBC}\tr \big[ \bar\Psi_B \g^\m \e^I( X_C X^A D_\m X_A -
D_\m X_A X^A X_C + X_A D_\m X^A X_C - X_C D_\m X^A X_A) \big].
\end{equation}
The terms of the indicated structure that arise from varying $L_{\rm
4a}$ are
$$
-2i \e^{ABCD}\tr(\d\bar\Psi_D X_A \Psi_B X_C) = -2i
\e^{ABCD}\G^I_{DE}\tr(\bar\Psi_B \g^\m \e^I X_C D_\m X^E X_A)
$$
\begin{equation}
= i\d^{ABC}_{EFG} \tilde\G^{IFG}\tr(\bar\Psi_B \g^\m \e^I X_C D_\m
X^E X_A)
\end{equation}
$$
= 2i \tilde\G^{IBC}\tr( \bar\Psi_B \g^\m \e^I X_C D_\m X^A X_A +
\bar\Psi_C \g^\m \e^I X_A D_\m X^A X_B + \bar\Psi_A \g^\m \e^I X_B
D_\m X^A X_C).
$$
The terms of the indicated structure that arise from varying $L_{\rm
4b}$ are
$$
i\tr(\d\bar\Psi^B \Psi_B X_A X^A) - i\tr(\bar\Psi_B \d\Psi^B X^A
X_A)
$$
\begin{equation}
= i \tilde\G^{IBC}\tr\big[ \bar\Psi_B \g^\m \e^I (D_\m X_C X^A X_A -
X_A X^A D_\m X_C) \big].
\end{equation}
The terms of the indicated structure that arise from varying $L_{\rm
4c}$ are
$$
2i\tr(\bar\Psi_A \d\Psi^B X^A X_B) - 2i\tr(\d\bar\Psi^B \Psi_A X_B
X^A)
$$
\begin{equation}
= 2i \tilde\G^{IBC}\tr\big[ \bar\Psi_A \g^\m \e^I (X_B X^AD_\m X_C +
D_\m X_B X^A X_C) \big].
\end{equation}
Adding these up, we obtain
$$
2i\tilde\G^{IBC}\tr\big[\bar\Psi_A \g^\m \e^I D_\m(X_B X^A X_C)
\big]
$$
\begin{equation}
+i\tilde\G^{IBC}\tr\Big[\bar\Psi_B \g^\m \e^I \big(D_\m(X_C X^A X_A)
- D_\m(X_A X^A X_C)\big)\Big].
\end{equation}
Thus, this can cancel against a variation of the spinor field in the
spinor kinetic term for the choice
\begin{equation}
\d_3 \Psi^A = \tilde\G^{IAB} \e^I(X_C X^C X_B - X_B X^C X_C) -2
\tilde\G^{IBC} \e^IX_B X^A X_C.
\end{equation}

\subsubsection*{Determination of $V = -L_6$}

The next step is to determine $L_6$ by requiring that its $\d X$
variation cancels against the $\d_3 \Psi$ variation of $L_4$. A key
identity in the analysis is
\begin{equation}
\Gamma_{AB}^{I}\tilde{\Gamma}^{ICD} = -2\delta_{AB}^{CD}.
\label{eq:1}
\end{equation}
This is verified by showing that the two sides agree when contracted
with $\d^B_C$ as well as with $(\tilde\G^J \G^K - \tilde\G^K
\G^J)^B{}_C$. Since these are 16 linearly independent $4\times 4$
matrices, this constitutes a complete proof.

The supersymmetry variation of $L_{4}$, keeping all terms containing
$\Psi^A$ but not $\Psi_A$ (since the $\Psi_A$ terms work in the same
way) is
\[
\delta
L_{4}=-2i\epsilon_{ABCD}\tr\Big(\delta_{3}\bar{\Psi}^{A}X^{B}\Psi^{C}X^{D}\Big)
\]
\begin{equation}
+i\tr\Big(\delta_{3}\bar{\Psi}_{A}\big(X_{B}X^{B}\Psi^{A}-\Psi^{A}X^{B}X_{B}
+2\Psi^{B}X^{A}X_{B}-2X_{B}X^{A}\Psi^{B}\big)\Big),
\end{equation}
where, as derived previously,
\begin{equation}
\delta_{3}\bar{\Psi}^{A}=\Gamma_{HK}^{I}
\left[\frac{1}{2}\epsilon^{ACHK}\left(X_{D}X^{D}X_{C}-X_{C}X^{D}X_{D}\right)
-\epsilon^{FGHK}X_{F}X^{A}X_{G}\right]\bar{\epsilon}^{I},
\end{equation}
\begin{equation}
\delta_{3}\bar{\Psi}_{A}=\left[-\Gamma_{AC}^{I}\left(X^{C}X_{D}X^{D}
-X^{D}X_{D}X^{C}\right)+2\Gamma_{HK}^{I}X^{K}X_{A}X^{H}\right]\bar{\epsilon}^{I}.
\end{equation}
Expanding $\delta L_{4}$ is straightforward algebra and gives
\[
\tr\Big( 3X^{A}\delta X_{A}X^{B}X_{B}X^{C}X_{C}+3\delta
X_{A}X^{A}X_{B}X^{B}X_{C}X^{C}
\]
\[
-2X^{A}\delta X_{B}X^{B}X_{A}X^{C}X_{C}-2X^{A}X_{B}X^{B}\delta
X_{A}X^{C}X_{C}-2X^{A}X_{B}X^{B}X_{A}X^{C}\delta X_{C}
\]
\begin{equation}
+4i\Gamma_{HK}^{I}\bar{\epsilon}^{I}\Psi^{A}
\left[X^{H}X_{A}X^{B}X_{B}X^{K}+X^{B}X_{B}X^{H}X_{A}X^{K}
+X^{H}X_{B}X^{K}X_{A}X^{B}\right.\label{eq:4}
\end{equation}
\[
\left.-X^{H}X_{B}X^{B}X_{A}X^{K}-X^{B}X_{A}X^{H}X_{B}X^{K}
-X^{H}X_{A}X^{K}X_{B}X^{B}\right]
\]
\[
+2i\epsilon_{ABCD}\epsilon^{FGHK}\Gamma_{HK}^{I}
\bar{\epsilon}^{I}\Psi^{A}X^{B}X_{F}X^{C}X_{G}X^{D}\Big).
\]
The first two lines can be reproduced by varying
\begin{equation}
V_1 = \tr\Big(X^{A}X_{A}X^{B}X_{B}X^{C}X_{C}
+X_{A}X^{A}X_{B}X^{B}X_{C}X^{C}
-2X^{A}X_{B}X^{B}X_{A}X^{C}X_{C}\Big).\label{eq:7}
\end{equation}
The last line cancels the third and fourth lines and contributes
additional terms to $V_1$, as we will now show. For this purpose,
the following identity is useful:
\newpage
\[
2\epsilon_{ABCD}\epsilon^{FGHK}\Gamma_{HK}^{I}
=\epsilon_{LBCD}\epsilon^{FGHK}\Gamma_{HK}^{J}
\left(2\delta^{IJ}\delta_{A}^{L}\right)
\]
\begin{equation}
=\epsilon_{LBCD}\epsilon^{FGHK}\Gamma_{HK}^{J}
\left(\Gamma_{AM}^{I}\tilde{\Gamma}^{JML}
+\Gamma_{AM}^{J}\tilde{\Gamma}^{IML}\right)
\end{equation}
\[
=4\delta_{BCD}^{FGM}\Gamma_{AM}^{I}+2\left(\delta_{BCD}^{GPQ}\delta_{A}^{F}
-\delta_{BCD}^{FPQ}\delta_{A}^{G}\right)\Gamma_{PQ}^{I},
\]
where we have used (\ref{eq:1}) to go from the second line to the
third line. Plugging this identity into the last line of
(\ref{eq:4}) gives
\[
\tr\Big(-4\delta_{BCD}^{FGM}\delta X_{M}X^{B}X_{F}X^{C}X_{G}X^{D}
\]
\begin{equation}
+2i\Gamma_{HK}^{I}
\bar{\epsilon}^{I}\Psi^{A}\left(\delta_{BCD}^{GHK}\delta_{A}^{F}
-\delta_{BCD}^{FHK}\delta_{A}^{G}\right)
X^{B}X_{F}X^{C}X_{G}X^{D}\Big).\label{eq:8}
\end{equation}
Expanding the first term in (\ref{eq:8}) gives
\[
4\tr\left[-X^{D}\delta X_{D}X^{F}X_{F}X^{G}X_{G}-\delta
X_{B}X^{B}X_{C}X^{C}X_{D}X^{D}-\delta
X_{C}X^{G}X_{D}X^{C}X_{G}X^{D}\right.
\]
\begin{equation}
\left.+\delta X_{C}X^{F}X_{F}X^{C}X_{D}X^{D}+\delta
X_{B}X^{B}X_{D}X^{G}X_{G}X^{D}+\delta
X_{D}X^{G}X_{C}X^{C}X_{G}X^{D}\right],
\end{equation}
which also comes from varying
$$
V_2 = \tr\Big(-\frac{4}{3}X^{A}X_{A}X^{B}X_{B}X^{C}X_{C}
-\frac{4}{3}X_{A}X^{A}X_{B}X^{B}X_{C}X^{C}
$$
\begin{equation}
-\frac{4}{3}X_{A}X^{B}X_{C}X^{A}X_{B}X^{C}
+4X^{A}X_{B}X^{B}X_{A}X^{C}X_{C}\Big).
\end{equation}
Adding this potential to Eq.~(\ref{eq:7}) gives the total potential
\[
V = -\frac{1}{3}\tr\Big[X^{A}X_{A}X^{B}X_{B}X^{C}X_{C}
+X_{A}X^{A}X_{B}X^{B}X_{C}X^{C}
\]
\begin{equation}
+4X_{A}X^{B}X_{C}X^{A}X_{B}X^{C}
-6X^{A}X_{B}X^{B}X_{A}X^{C}X_{C}\Big] .
\end{equation}
Furthermore, straightforward algebra shows that the second term in
Eq.~(\ref{eq:8}) precisely cancels the terms in the third and fourth
lines of Eq.~(\ref{eq:4}). So we conclude that the variation of
$L_{4}$ is completely canceled by varying $-V$. This expression
agrees with the potential obtained in
\cite{Aharony:2008ug,Benna:2008zy}.

It is also interesting to note that $V$ is proportional to the trace
of the absolute square of the $X^{3}$ expression that appears in
$\delta_{3}\Psi$. Specifically,
\begin{equation}
V = \frac{1}{6}\tr(N^{IA} N^I_A),
\end{equation}
which is straightforward to verify using Eq.~(\ref{eq:1}).

\subsubsection*{Conserved Supersymmetry Current}

The conserved supersymmetry current of the $U(N) \times U(N)$
theory, generalizing the expression given earlier for the $U(1)
\times U(1)$ theory, is
\begin{equation}
Q_\m^I = \tr\Big( M^I_A  \g_\m \Psi^A\Big) + \tr\Big( M^{IA}  \g_\m
\Psi_A \Big).
\end{equation}
Here
\begin{equation}
M^I_A = -\G^I_{AB}\g\cdot D X^B + N^I_A
\end{equation}
and
\begin{equation}
M^{IA} = \tilde\G^{IAB}\g\cdot D X_B + N^{IA}
\end{equation}
are quantities that appear in the supersymmetry variations of the
spinor fields $\bar\Psi_A$ and $\bar\Psi^A$, respectively. The
quantity $N^I_A$ and its adjoint $N^{IA}$ were defined in Eqs.
(\ref{NsupA}) and (\ref{NsubA}). The verification that this current
is conserved as a consequence of the equations of motion is rather
tedious. In any case, it would be redundant, since it is equivalent
to the verification of the supersymmetry of the action, which we
have just carried out.

\subsubsection*{Conformal Supersymmetry}

In the $U(1)\times U(1)$ case, we found that the conformal
supersymmetries can be described by replacing $\e^I$ in the
Poincar\'e supersymmetries by $\g\cdot x\, \eta^I$ and by adding an
additional term to the spinor field transformations
\begin{equation}
\d'\Psi_A = \G^I_{AB}  X^B \eta^I
\end{equation}
and its adjoint. Let us now verify that the same rule continues to
work for $N>1$. Most terms cancel as a consequence of the Poincar\'e
supersymmetry. The remaining ones that need to cancel separately are
those that arise from the derivative in $i\bar\Psi_A \g \cdot
D\d\Psi^A$ acting on the explicit $x^\m$ in the $\eta^I$
transformation. This gives
\begin{equation}
i\bar\Psi_A\Big[\tilde\G^{IAB}(\g\cdot D X_B +3X_C X^C X_B - 3X_B
X^C X_C) -6 \tilde\G^{IBC} X_B X^A X_C\Big]\eta^I.
\end{equation}
The first term in this expression is canceled by the $\d'\Psi^A$
variation of the spinor kinetic term. The remaining terms need to
cancel against the $\d' \Psi$ variation of $L_4$. The relevant terms
that arise in this way are
$$
2i\e^{ABCD} \tr(\d'\bar\Psi_A  X_B \Psi_C  X_D) + i \tr(
\d'\bar\Psi^A \Psi_A X_B X^B) - i\tr( \bar\Psi_A \d'\Psi^A X^B X_B)
$$
\begin{equation}
+ 2i\tr( \bar\Psi_A \d'\Psi^B X^A X_B) - 2i\tr(\d' \bar\Psi^B \Psi_A
X_B X^A).
\end{equation}
By manipulations similar to those described previously, the first
term in this expression can be recast in the form
\begin{equation}
2i\tilde\G^{IBC}\tr(\bar\Psi_A X_B X^A X_C +\bar\Psi_B X_C X^A X_A
+\bar\Psi_C X_A X^A X_B)\eta^I.
\end{equation}
Combining this with the other four terms leaves
\begin{equation}
i\bar\Psi_A\Big[\tilde\G^{IAB}(-3X_C X^C X_B + 3X_B X^C X_C) +6
\tilde\G^{IBC} X_B X^A X_C\Big]\eta^I.
\end{equation}
This provides the desired cancellation, which proves that the theory
has conformal supersymmetry.

Taken together with the ${\cal N}=6$ Poincar\'e supersymmetry, the
conformal supersymmetry implies that the theory has the full
$OSp(6|4)$ superconformal symmetry. Even though this result is
necessary for a dual AdS interpretation, it was not at all obvious
that this symmetry would hold. After all, it is not a logical
consequence of the other symmetries that have been verified.

Accordingly, the conserved conformal supersymmetry currents in the
$U(N)\times U(N)$ theory are given by
\begin{equation}
S_\m^I = \g\cdot x \, Q_\m^I - \G^I_{AB} \tr\Big(X^B \g_\m \Psi^A
\Big)+ \tilde\G^{IAB} \tr\Big(X_B \g_\m \Psi_A\Big) .
\end{equation}
As a check on our analysis, let us compute the divergence. The
$DX^B$ terms cancel leaving
\begin{equation}
\pa^\m S_\m^I = \tr\Big(3N^I_A \Psi^A +3 N^{IA} \Psi_A
-\G^I_{AB}X^B\g\cdot D \Psi^A  + \tilde\G^{IAB}X_B\g\cdot D \Psi_A
\Big),
\end{equation}
where $ N^I_A $ and $ N^{IA}$ are as before. Using the spinor field
equations of motion (\ref{PsiEOMa}) and (\ref{PsiEOMb}) to eliminate
$\g\cdot D \Psi^A$ and $\g\cdot D \Psi_A$, the terms in $\pa^\m
S_\m^I$ that involve $\Psi^A$ are \pagebreak
$$
3\tr\Big(N^I_A \Psi^A\Big)+ 2\e_{ACDE} \tilde\G^{IAB}\tr\Big(X_B X^C
\Psi^D X^E \Big)
$$
\begin{equation}
-\G^I_{AB}\tr\Big(X^B[-X_C X^C \Psi^A + \Psi^A X^C X_C -2 \Psi^C X^A
X_C +2 X_C X^A \Psi^C ]\Big).
\end{equation}
A short calculation, similar to previous ones, shows that this
vanishes.


\newpage

\begin{thebibliography}{99}

\bibitem{Aharony:2008ug}
  O.~Aharony, O.~Bergman, D.~L.~Jafferis and J.~Maldacena,
  ``$N=6$ Superconformal Chern-Simons-Matter Theories, M2-branes and Their
  Gravity Duals,''
  arXiv:0806.1218 [hep-th].

\bibitem{Benna:2008zy}
  M.~Benna, I.~Klebanov, T.~Klose and M.~Smedback,
  ``Superconformal Chern-Simons Theories and AdS${}_4$/CFT${}_3$ Correspondence,''
  arXiv:0806.1519 [hep-th].

\bibitem{Bhattacharya:2008bj}
  J.~Bhattacharya and S.~Minwalla,
  ``Superconformal Indices for ${\cal N}=6$ Chern Simons Theories,''
  arXiv:0806.3251 [hep-th].

\bibitem{Nishioka:2008gz}
  T.~Nishioka and T.~Takayanagi,
  ``On Type IIA Penrose Limit and $N=6$ Chern-Simons Theories,''
  arXiv:0806.3391 [hep-th].

\bibitem{Honma:2008jd}
  Y.~Honma, S.~Iso, Y.~Sumitomo and S.~Zhang,
  ``Scaling Limit of $N=6$ superconformal Chern-Simons Theories and Lorentzian
  Bagger-Lambert Theories,''
  arXiv:0806.3498 [hep-th].

\bibitem{Imamura:2008nn}
  Y.~Imamura and K.~Kimura,
  ``Coulomb Branch of Generalized ABJM Models,''
  arXiv:0806.3727 [hep-th].

\bibitem{Minahan:2008hf}
  J.~A.~Minahan and K.~Zarembo,
  ``The Bethe Ansatz for Superconformal Chern-Simons,''
  arXiv:0806.3951 [hep-th].

\bibitem{Armoni:2008kr}
  A.~Armoni and A.~Naqvi,
  ``A Non-Supersymmetric Large-N 3D CFT and Its Gravity Dual,''
  arXiv:0806.4068 [hep-th].

\bibitem{Gaiotto:2008cg}
  D.~Gaiotto, S.~Giombi and X.~Yin,
  ``Spin Chains in $N=6$ Superconformal Chern-Simons-Matter Theory,''
  arXiv:0806.4589 [hep-th].

\bibitem{Grignani:2008is}
  G.~Grignani, T.~Harmark and M.~Orselli,
  ``The SU(2) x SU(2) Sector in the String Dual of $N=6$ Superconformal
  Chern-Simons Theory,''
  arXiv:0806.4959 [hep-th].

\bibitem{Hosomichi:2008jb}
  K.~Hosomichi, K.~M.~Lee, S.~Lee, S.~Lee and J.~Park,
  ``$N=5,6$ Superconformal Chern--Simons Theories and M2-branes on Orbifolds,''
  arXiv:0806.4977 [hep-th].

\bibitem{Hanany:2008qc}
  A.~Hanany, N.~Mekareeya and A.~Zaffaroni,
  ``Partition Functions for Membrane Theories,''
  arXiv:0806.4212 [hep-th].

\bibitem{Bagger:2008se}
  J.~Bagger and N.~Lambert,
  ``Three-Algebras and N=6 Chern--Simons Gauge Theories,''
  arXiv:0807.0163 [hep-th].

\bibitem{Terashima:2008sy}
  S.~Terashima,
  ``On M5-branes in $N=6$ Membrane Action,''
  arXiv:0807.0197 [hep-th].

\bibitem{Grignani:2008te}
  G.~Grignani, T.~Harmark, M.~Orselli and G.~W.~Semenoff,
  ``Finite Size Giant Magnons in the String Dual of $N=6$ Superconformal
  Chern-Simons Theory,''
  arXiv:0807.0205 [hep-th].

\bibitem{Schwarz:2004yj}
  J.~H.~Schwarz,
  ``Superconformal Chern--Simons Theories,''
  JHEP {\bf 0411}, 078 (2004)
  [arXiv:hep-th/0411077].

\bibitem{Basu:2004ed}
  A.~Basu and J.~A.~Harvey,
  ``The M2--M5 Brane System and a Generalized Nahm's Equation,''
  Nucl.\ Phys.\  B {\bf 713}, 136 (2005)
  [arXiv:hep-th/0412310].

\bibitem{Bagger:2006sk}
  J.~Bagger and N.~Lambert,
  ``Modeling Multiple M2's,''
  Phys.\ Rev.\  D {\bf 75}, 045020 (2007)
  [arXiv:hep-th/0611108].

\bibitem{Bagger:2007jr}
  J.~Bagger and N.~Lambert,
  ``Gauge Symmetry and Supersymmetry of Multiple M2-Branes,''
  Phys.\ Rev.\  D {\bf 77}, 065008 (2008)
  [arXiv:0711.0955 [hep-th]].

\bibitem{Bagger:2007vi}
  J.~Bagger and N.~Lambert,
  ``Comments on Multiple M2-branes,''
  JHEP {\bf 0802}, 105 (2008)
  [arXiv:0712.3738 [hep-th]].

\bibitem{Gustavsson:2007vu}
  A.~Gustavsson,
  ``Algebraic Structures on Parallel M2-Branes,''
  arXiv:0709.1260 [hep-th].

\bibitem{Gustavsson:2008dy}
  A.~Gustavsson,
  ``Selfdual Strings and Loop Space Nahm Equations,''
  JHEP {\bf 0804}, 083 (2008)
  [arXiv:0802.3456 [hep-th]].

\bibitem{Bandres:2008vf}
  M.~A.~Bandres, A.~E.~Lipstein and J.~H.~Schwarz,
  ``$N = 8$ Superconformal Chern--Simons Theories,''
  JHEP {\bf 0805}, 025 (2008)
  [arXiv:0803.3242 [hep-th]].

\bibitem{VanRaamsdonk:2008ft}
  M.~Van Raamsdonk,
  ``Comments on the Bagger-Lambert Theory and Multiple M2-branes,''
  JHEP {\bf 0805}, 105 (2008)
  [arXiv:0803.3803 [hep-th]].

\bibitem{Papadopoulos:2008sk}
  G.~Papadopoulos,
  ``M2-branes, 3-Lie Algebras and Plucker Relations,''
  JHEP {\bf 0805}, 054 (2008)
  [arXiv:0804.2662 [hep-th]].

\bibitem{Gauntlett:2008uf}
  J.~P.~Gauntlett and J.~B.~Gutowski,
  ``Constraining Maximally Supersymmetric Membrane Actions,''
  arXiv:0804.3078 [hep-th].

\bibitem{Gomis:2008uv}
  J.~Gomis, G.~Milanesi and J.~G.~Russo,
  ``Bagger-Lambert Theory for General Lie Algebras,''
  arXiv:0805.1012 [hep-th].

\bibitem{Benvenuti:2008bt}
  S.~Benvenuti, D.~Rodriguez-Gomez, E.~Tonni and H.~Verlinde,
  ``N=8 Superconformal Gauge Theories and M2 Branes,''
  arXiv:0805.1087 [hep-th].

\bibitem{Ho:2008ei}
  P.~M.~Ho, Y.~Imamura and Y.~Matsuo,
  ``M2 to D2 Revisited,''
  arXiv:0805.1202 [hep-th].

\bibitem{Bandres:2008kj}
  M.~A.~Bandres, A.~E.~Lipstein and J.~H.~Schwarz,
  ``Ghost-Free Superconformal Action for Multiple M2-Branes,''
  arXiv:0806.0054 [hep-th].

\bibitem{Gomis:2008be}
  J.~Gomis, D.~Rodriguez-Gomez, M.~Van Raamsdonk and H.~Verlinde,
  ``The Superconformal Gauge Theory on M2-Branes,''
  arXiv:0806.0738 [hep-th].

\bibitem{Ezhuthachan:2008ch}
  B.~Ezhuthachan, S.~Mukhi and C.~Papageorgakis,
  ``D2 to D2,''
  arXiv:0806.1639 [hep-th].

\bibitem{Gaiotto:2008sd}
  D.~Gaiotto and E.~Witten,
  ``Janus Configurations, Chern-Simons Couplings, and the Theta-Angle in
  $N=4$ Super Yang-Mills Theory,''
  arXiv:0804.2907 [hep-th].

\bibitem{Hosomichi:2008jd}
  K.~Hosomichi, K.~M.~Lee, S.~Lee, S.~Lee and J.~Park,
  ``$N=4$ Superconformal Chern-Simons Theories with Hyper and Twisted Hyper
  Multiplets,''
  arXiv:0805.3662 [hep-th].

\bibitem{Cattaneo:1995tw}
  A.~S.~Cattaneo, P.~Cotta-Ramusino, J.~Frohlich and M.~Martellini,
  ``Topological BF Theories in Three Dimensions and Four Dimensions,''
  J.\ Math.\ Phys.\  {\bf 36}, 6137 (1995)
  [arXiv:hep-th/9505027].

\bibitem{Lambert:2008et}
  N.~Lambert and D.~Tong,
  ``Membranes on an Orbifold,''
  arXiv:0804.1114 [hep-th].

\bibitem{Distler:2008mk}
  J.~Distler, S.~Mukhi, C.~Papageorgakis and M.~Van Raamsdonk,
  ``M2-branes on M-folds,''
  JHEP {\bf 0805}, 038 (2008)
  [arXiv:0804.1256 [hep-th]].


\end{thebibliography}
\end{document}